\documentclass{article}
\newcommand{\grad}{\vec{\nabla}}
\newcommand{\bound}{\Gamma(\Omega)}

\title{Generalization of Hamilton-Jacobi method and its consequences in classical, relativistic, and quantum mechanics}
\author{O. Chavoya-Aceves\\Camelback High School\\
4612 North 28th Street
\\Phoenix, AZ, 85016, USA}
\begin{document}
\maketitle
\begin{abstract}
The Hamilton-Jacobi method is generalized in classical and
relativistic mechanics. The implications in quantum mechanics are
discussed in the case of Klein-Gordon equation. We find that the
wave functions of Klein-Gordon theory can be soundly interpreted
as describing the motion of an ensemble of particles that move
under the action of the electromagnetic field alone, without
quantum potentials, hidden uninterpreted variables, or zero point
fields.

\noindent\textbf{pacs 03.20.+i, 03.20.+p, 03.65.-w}

\end{abstract}

\section{Introduction}
In previous papers (\cite{CHAVOYAI}, \cite{CHAVOYAII}) we have
considered the possibility of reinterpreting Klein-Gordon and
Dirac's equations, as describing ensembles of particles moving
under the action of the electromagnetic field, where the number of
particles is not locally conserved. In order to do that for the
Klein-Gordon equation, on the basis of some general electrodynamic
considerations, we proved that it was sound to assume that the
field of kinetic four-momentum \[
p^{\mu}=\left(\frac{mc}{\sqrt{1-\frac{v^2}{c^2}}},\frac{m\vec{v}}{\sqrt{1-\frac{v^2}{c^2}}},\right)
\] is given by
\begin{equation}\label{campo de velocidades klein gordon}
    p_{\mu}=-\partial_{\mu} S - \partial_{\mu} \Phi -
    \frac{q}{c}A_u;
\end{equation}
where $S$ is the phase of the wave function; $A_{\mu}$ is the
electrodynamic four-potential; and $\Phi$ is a function of
space-time coordinates, whose nature we left completely
undetermined. To deal with Dirac's equation we used a similar
approach.

In this paper we determine the nature of the function $\Phi$.
Actually, we prove that the principle of relativity requires that
we consider a general four-vector $\omega_{\mu}$, instead of the
gradient of a potential. For this we start with an analysis of
Helmholtz's theorem on the representation of vectors fields and
its extensions to $n$ dimensional spaces, in particular to space
time\cite{WOODSIDE}. In the first section we derive Helmholtz
theorem:
\begin{equation}\label{expresion del teorema de helmholtz}
  \vec{f}=\grad \phi+\grad\times\vec{\lambda},
\end{equation}
following a variational approach. This permits us to prove that if
the potentials are chosen in a manner that
$(\grad\phi-\vec{f})\cdot\hat{n}=0$ at the boundary of the region
under consideration, then $\grad\phi$ is the best approximation to
$\vec{f}$, in the sense of the quadratic norm---which we think is
an original result.

Further we apply those conclusions to perform an analysis of the
physical meaning of Hamilton-Jacobi theory. Taking it out of the
mathematical realm of canonical transformations, allows us to find
its necessary generalizations in classical and relativistic
mechanics, laying thus a foundation for a complete explanation of
the field $\omega_{\mu}$. In the second section we work out a
complete example of this extension of Hamilton-Jacobi theory in a
very simple classical situation, just to show how it works, as
well as to prove that there are very simple and common dynamical
scenarios that are being disregarded by the present theory and,
therefore, that our extension is not only a theoretical
speculation, but a necessary step to extend the scope and strength
of a fundamental part of classical mechanics.

We consider a flux that corresponds to an infinite ensemble of
particles with the initial velocities:
\begin{equation}\label{condiciones iniciales del flujo}
    \vec{v}(\vec{r},0)=\vec{\omega}\times\vec{r},
\end{equation}
where $\vec{\omega}$ is a constant vector. We prove that this flux
is invertible, in the sense that we can obtain the initial
position of the particle that passes through the point $\vec{x}$
at time $t$. However this flux cannot be derived from a potential
in the form
\begin{equation}\label{gradiente de la accion}
    \vec{v}=\frac{\grad S}{m},
\end{equation}
because
\[
\grad\times\vec{v}\ne \vec{0}.
\]

The generalization of Hamilton-Jacobi theory is even more
necessary in the case of special relativity because---as we will
prove---any field of four-velocities that describes the motion of
an ensemble of particles has a vorticity, at least it describes a
flux of free particles. Furthermore: all we have is to consider a
relativistic ensemble of free particles, where the initial
momentum is defined as
\begin{equation}\label{condiciones iniciales del flujo}
    \vec{p}(\vec{r},0)=\vec{\omega}\times\vec{r},
\end{equation}
for a particular observer, to have a system whose evolution cannot
be described by the Hamilton-Jacobi theory as it is presented in
\cite[pp. 24-29]{LANDAU}. In some way we follow Bohr when we state
that if there is a reformulation of Hamilton-Jacobi theory that
widens its range of applicability it should be incorporated into
the body of mechanical knowledge, no matter if it is or  not
compatible with the supremacy of Hamilton's principal function.
\begin{quote}
  In my opinion, there could be no other way to deem a logically
  consistent mathematical formalism as inadequate than by demonstrating the departure of
  its consequences from experience \emph{or by proving that its predictions did not exhaust the possibilities of
  observation}\cite[pp. 200-241]{BOHR}.
\end{quote}

We consider this kind of fluxes, because they appear in the
classical limit of quantum mechanics:

\begin{quote}
 In the classical approximation, $\psi$ describes a fluid of non-interacting classical
 particles of mass $m$, and subject to the potential $V(\vec{r})$:
 the density and current density of this fluid at each point of
 space are respectively equal to the probability density $P$ and the probability current $\vec{j}$ of
 the quantum particle at that point.

 Indeed, since the continuity equation of this fluid is satisfied
 [eq. (VI.19)], it suffices to show that the velocity field\[
 \vec{v}=\frac{\vec{j}}{P}=\frac{\grad S}{m}
 \]
 of this fluid, actually follows the law of motion of the
 classical fluid in question.
 \cite[p. 223]{MESSIAH}.
\end{quote}

In this way, this paper demonstrates that there are classical
scenarios, as the one described by (\ref{condiciones iniciales del
flujo}) that cannot be reproduced from quantum mechanics in the
classical limit.

In the last section we use our generalization of the
Hamilton-Jacobi theory to prove---as a mathematical fact, because
our proof is logical, not ontological---that the wave functions of
Klein-Gordon theory can be soundly interpreted as describing the
motion of ensembles of particles under the action of the
electromagnetic field, alone, without any quantum potentials,
uninterpreted hidden variables, or zero point fields. What Bohm
called the \emph{quantum potential} corresponds in part to kinetic
energy and in part to reaction forces associated to local
creation/annihilation processes. In this form, we address the only
objection that Einstein raised against the de Broglie-Bohm
interpretation of quantum mechanics\cite{BOHM}\cite[pp.
33-40]{EINSTEIN}: that, according to Bohm's theory, in the
stationary states of particles inside a box with rigid walls, the
particles had to be at rest.

The reader should realize that to say that ``the wave functions of
Klein-Gordon theory can be soundly interpreted as describing
ensembles of particles under the action of the electromagnetic
field...'' is not the same that to say ``the wave functions of
Klein-Gordon theory describe ensembles of particles under the
action of the electromagnetic field...'' Therefore, we are not
making any ontological assertions in this paper, which keeps us
out of the, otherwise fascinating, realm of philosophical
speculation.

We do not say, like Einstein-Podolsky-Rosen, that quantum
mechanics is not a complete theory of motion\cite{EPR}. Instead,
we say that the set of terms that specify the wave function in
Madelung's representation, $\{\rho,S\}$\cite{MADELUNG}, can be
complemented introducing a four vector $\omega_i$, in such manner
that the same phenomena can be explained using the classical
concept of particles that move along well defined trajectories. In
addition, the logical necessity of introducing this four-vector
will be mathematically proved beyond any doubt on the basis of
Helmholtz theorem.

This four-vector appears in the equations as the potential of a
kind of electromagnetic field, that does not produce any Lorentz
force on the particles of the ensemble under examination. The
electric force exactly cancels the magnetic force. However there
is not any \emph{a priori} condition on the divergence of the
corresponding Faraday tensor. Therefore this field can have an
associated \emph{virtual electric current}.

\section{The Helmholtz's Theorem in Three Dimensions}
To find the best approximation of a vector field $\vec{f}$ by
means of the gradient of a potential in a region of space we can
follow a variational approach:

\noindent\textbf{ Determine the extreme of the functional:}
\begin{equation}\label{problema ejemplo}
    F=\frac{1}{2}\int_{\Omega}(\grad\phi-\vec{f})^2dV.
\end{equation}

The first variation is:
\begin{equation}\label{primera variacion problema ejemplo}
\delta F = \int_{\Omega}\delta\grad\phi\cdot(\grad\phi-\vec{f})
dV=
\end{equation}
\[\int_{\Omega}\grad\cdot[\delta\phi(\grad\phi-\vec{f})]dV-\int_{\Omega}\delta\phi\grad\cdot(\grad\phi-\vec{f})dV.\]

Making use of Gauss theorem, the first of the last integrals is
replaced by an integral on $\Omega$'s boundary($\bound$).
Therefore,
\begin{equation}\label{desarrollo de la primera variacion}
    \delta F=\int_{\bound}\delta\phi(\grad\phi-\vec{f})\cdot
    d\vec{S}-\int_{\Omega}\delta\phi\grad\cdot(\grad\phi-\vec{f})dV
\end{equation}

We do not fix the values of $\phi$ on $\bound$, so that, from the
condition $\delta F=0$, we can get a differential equation and a
set of boundary conditions:
\begin{equation}\label{ecuacion de poisson}
    \grad^2\phi=\grad\cdot\vec{f},
\end{equation}
\begin{equation}\label{condiciones de frontera}
    \forall \vec{x}\in \bound :\
    (\grad\phi-\vec{f})\cdot\hat{n}(x)=0.
\end{equation}
(Where $\hat{n}(\vec{x})$ is the field of unitary normals on
$\bound$.)

From the theory of harmonic functions we know the last problem has
a unique solution which we will not discuss further. The field
$\vec{f}$ is thus expressed as the sum of two fields:
\begin{equation}\label{descomposicion de helmholtz}
    \vec{f}=\grad\phi+\vec{t},
\end{equation}
where $\vec{t}$ is a solenoidal field, such that
\begin{equation}\label{rotacional}
    \grad\times\vec{f}=\grad\times\vec{t},
\end{equation}
and
\[
\frac{1}{2}\int_{\Omega}\vec{t}^2 dV
\]
has the minimum value compatible with equation (\ref{rotacional}).

Therefore, the  field $\vec{t}$ is a solution for another
variational problem:
\begin{equation}\label{primer problema}
    \delta\int_\Omega\left[\frac{1}{2}\vec{t}^2 +
    \vec{\lambda}\cdot \grad\times(\vec{f}-\vec{t})\right]dV,
\end{equation}
where the components of $\vec{\lambda}$ are Lagrange's
multipliers.

The variation with respect to $\vec{\lambda}$ leads to the
condition $\grad\times\vec{t}=\grad\times\vec{f}$, which we
already know.

The variation with respect to $\vec{t}$ leads to:
\begin{equation}\label{primera variacion primer problema}
    \int_\Omega \left[\vec{t}\cdot\delta\vec{t} %
    -\vec{\lambda}\cdot
    \grad\times\delta\vec{t}\ \right]dV=0.
\end{equation}

Using the identity
\[
\grad\cdot(\vec{a}\times\vec{b})=\vec{b}\cdot\grad\times\vec{a}-\vec{a}\cdot\grad\times\vec{b}
\]
to obtain the substitution
\[
-\vec\lambda\cdot\grad\times\delta\vec
t=\grad\cdot(\vec\lambda\times\delta\vec t)-\delta\vec t\cdot
\grad\times\vec\lambda,
\]
 equation (\ref{primera variacion
primer problema}) can be written as:
\begin{equation}\label{desarrollo primera variacion primer problema}
\int_\Omega \left[(\vec{t}-\grad\times\vec{\lambda})\cdot\delta\vec{t}%
    +\grad\cdot(\vec{\lambda}\times\delta\vec{t})\right]dV=
\end{equation}
\[
\int_\Omega
(\vec{t}-\grad\times\vec{\lambda})\cdot\delta\vec{t}dV%
+\int_{\bound} (\hat{n}\times\vec{\lambda})\cdot\delta\vec{t}dS=0.
\]

From this we get the differential equation:
\begin{equation}\label{condicion de extremo primer problema}
    \vec{t}=\grad\times\vec{\lambda}
\end{equation}

From equation (\ref{rotacional}) we get:
\begin{equation}\label{ecuacion vectorial de poisson}
    \grad\times(\grad\times\vec{\lambda})=\grad\times\vec{f}.
\end{equation}

We can choose $\vec{\lambda}$ so that $\grad\cdot\vec{\lambda}=0$.
For this all we have to do is the substitution:
\[
\vec\lambda\rightarrow\vec\lambda+\grad \psi,
\]
where $\psi$ is a particular solution of a Poisson equation
\[
\grad^2\psi=-\grad\cdot\vec\lambda.
\]
This transformation does not change the fundamental relation
(\ref{condicion de extremo primer problema}).

Equation (\ref{ecuacion vectorial de poisson}) is transformed
into:
\begin{equation}\label{ecuacion vectorial de poisson simplificada}
    \grad^2\vec{\lambda}=-\grad\times\vec{f}.
\end{equation}
with the extra and boundary conditions:
\begin{equation}\label{divergencia del potencial vectorial viz}
    \grad\cdot\vec{\lambda}=0
\end{equation}
\begin{equation}\label{condiciones de frontera para el potencial vectorial}
    \forall \vec x \in \bound (\grad\times\vec\lambda - \vec t)\times \hat n  =0
\end{equation}
Those relations univocally determine the field $\vec t$.

We have thus an addition to Helmholtz' theorem in the form:

\begin{quote}Every vector field $\vec f$ with continuous partial
derivatives inside a region $\Omega$ can be written in the form:
\[
\vec f=\grad \phi +\grad\times\vec \lambda,
\]
where $\vec\lambda$ is solenoidal and $\phi$ and $\vec\lambda$ are
solutions of the Poisson equations:
\[\grad^2\phi=\grad\cdot\vec
f,\] and \[\grad^2\vec\lambda=-\grad\times\vec f.\] If $\phi$ and
$\vec\lambda$ are chosen to meet the boundary condition
\[\forall \vec x \in \bound:\ (\grad\phi-\vec f)\cdot\hat n(\vec
x)=0,\] then $\grad \phi$ is the best approximation of the field
$\vec f$ as the gradient of a scalar function.
\end{quote}

Consider now a vector field $(f_1,\cdots,f_n)$ in a space of $n$
dimensions. As before, we find the solutions of a variational
problem $\delta F=0$, where
\begin{equation}\label{primer problema n dimensiones}
    F=\frac{1}{2}\int_{\Omega} (f_i-\frac{\partial \phi}{\partial
    x_i})(f_i-\frac{\partial \phi}{\partial
    x_i})dV,
\end{equation}
where $\Omega$ is a region of a $n$-dimensional space limited by a
hyper-surface $\bound$. (We are using the summation convention.)

The condition $\delta F=0$ leads to a set of corresponding
equations, analogous to (\ref{ecuacion de poisson},
\ref{condiciones de frontera}). As to the vector
$t_i=f_i-\frac{\partial \phi}{\partial x_i}$, we know it is
solenoidal, and that it is a solution of the variational problem
\begin{equation}\label{segundo problema n dimensiones}
    \delta \int_{\Omega} \frac{1}{2} t_it_i +
    \Lambda_{ij}\left(\frac{\partial }{\partial x_j}(t_i-f_i)-\frac{\partial}{\partial
    x_i}(t_j-f_j)\right)dV=0,
\end{equation}
where $\Lambda_{ij}$ is an antisymmetric set of Lagrange's
multipliers.

The extreme condition is easily found to be:
\begin{equation}\label{introduccion del tensor de vorticidad}
    t_i=\frac{\partial \Lambda _{ij}}{\partial x_j},
\end{equation}

Therefore, $t_i$ is the divergence of an antisymmetric second
order (cartesian) tensor.

Writing
\begin{equation}\label{introduccion del potencial tensorial}
    \Phi_{ij}=\Lambda_{ij}+\delta_{ij}\phi,
\end{equation}
we can see that
\begin{equation}\label{introduccion del potencial tensorial vis}
    f_i=\frac{\partial \Phi_{ij}}{\partial x_j},
\end{equation}
meaning that any vector field can be written as the divergence of
a second order (cartesian) tensor.


\section{Classical Mechanics}
Let $p_i(q_1,\cdots,q_n,t)$ be a vector field, defined in the
space-time of a mechanical system whose evolution is determined by
a Hamilton function \[H(q_1,\cdots,q_n,p_1,\cdots,p_n,t),\] so
that:
\begin{equation}\label{ecuaciones de hamilton}
    \dot q_i = \frac{\partial H}{\partial p_i}\ \dot
    p_i=-\frac{\partial H}{\partial q_i}.
\end{equation}
(Given that functions defined in Cartan's \emph{espace des
\'etats} \cite[174]{LANCZOS} will be projected into functions
defined in space-time, to avoid confusions, we will use symbols of
the kind
\begin{equation}\label{notacion}
    G_{\alpha\beta\cdots}
\end{equation}
to represent their derivatives. For example:
\[
\frac{\partial G}{\partial q_i}= G_{q_{i}}.
\]
To represent the derivatives with respect to the space-time
coordinates we will use the usual notation $ \frac{\partial
}{\partial q_i}, \frac{\partial }{\partial t} $.)

 We suppose the $p_i$ are the momenta of an infinite ensemble of particles, of the kind
used in Euler's hydro-kinematics. The particle that occupies the
position $q_i$ at time $t$, gets an increment of action in the
interval $dt$, which is given by:
\begin{equation}\label{incremento de la accion}
    dS=\sum p_idq_i-Hdt.
\end{equation}
The conditions under which this is relation is integrable are
\begin{equation}\label{condicion de zero vorticidad del impulso}
    \frac{\partial p_i}{\partial q_j}-\frac{\partial p_j}{\partial
    q_i}=0,
\end{equation}
and
\begin{equation}\label{segunda condicion de integrabilidad}
    \frac{\partial p_i}{\partial t}+\frac{\partial H}{\partial
    q_i}=0.
\end{equation}

The derivative of $p_i$ along the corresponding trajectory in
phase space is:
\begin{equation}\label{derivada material del impulso}
    \dot p_i=-H_{q_i}=\frac{\partial p_i}{\partial t}+\sum \dot q_j \frac{\partial p_i}{\partial
    q_j}.
\end{equation}
From this we get the equation:
\begin{equation}\label{derivada temporal del impulso}
    \frac{\partial p_i}{\partial t}=-H_{q_i}-\sum \dot q_j\frac{\partial p_i}{\partial
    q_j}.
\end{equation}
Also,
\begin{equation}\label{derivada espacial del hamiltoniano}
    \frac{\partial H}{\partial q_i}=H_{q_i}+\sum H_{p_j}\frac{\partial p_j}{\partial
    q_i}=H_{q_i}+\sum \dot q_j \frac{\partial p_j}{\partial
    q_i}.%
\end{equation}
Therefore:
\[
\frac{\partial p_i}{\partial t}+\frac{\partial H}{\partial
q_i}=\sum \dot q_j (\frac{\partial p_j}{\partial
q_i}-\frac{\partial p_i}{\partial q_j}).
\]

Thus, we get the important conclusion that the vorticity of
$(p_1,\cdots,p_n)$  is zero in the configuration space if and only
if the vorticity of $(-p_1,\cdots,-p_n,H)$ is zero in space-time.

According to equation (\ref{derivada temporal del impulso})
\begin{equation}\label{teorema de la vorticidad primera ecuacion}
    \frac{\partial^2 p_i}{\partial t\partial q_j}=-\frac{\partial H_{q_i}}{\partial
    q_j}-\sum \frac{\partial H_{p_k}}{\partial q_j}\frac{\partial p_i}{\partial
    q_k}-\sum H_{p_k}\frac{\partial^2 p_i }{\partial q_k q_j}.
\end{equation}
From this we can see that:
\begin{equation}\label{derivada material de la vorticidad}
    \frac{\partial }{\partial t}\left( \frac{\partial p_i}{\partial q_j}-\frac{\partial p_j}{\partial
    q_i}\right)+\sum\dot q_k\frac{\partial }{\partial q_k}\left( \frac{\partial p_i}{\partial q_j}-\frac{\partial p_j}{\partial
    q_i}\right)=
\end{equation}
\[
\frac{\partial H_{q_j}}{\partial q_i}-\frac{\partial H_{q_i}
}{\partial q_j}+\sum \frac{\partial H_{p_k}}{\partial
q_i}\frac{\partial p_j}{\partial q_k}-\frac{\partial
H_{p_k}}{\partial q_j}\frac{\partial p_i}{\partial q_k}.
\]

Considering the relations
\begin{equation}\label{teorema de la vorticidad segunda ecuacion}
    \frac{\partial H_{q_j}}{\partial q_i}-\frac{\partial H_{q_i}
}{\partial q_j}=\sum H_{q_jp_k}\frac{\partial p_k}{\partial
q_i}-H_{q_ip_k}\frac{\partial p_k}{\partial q_j},
\end{equation}
and
\begin{equation}\label{teorema de la vorticidad tercera ecuacion}
\sum \frac{\partial H_{p_k}}{\partial q_i}\frac{\partial
p_j}{\partial q_k}-\frac{\partial H_{p_k}}{\partial
q_j}\frac{\partial p_i}{\partial q_k}=
\end{equation}
\[
\sum\left(H_{p_kq_i}\frac{\partial p_j}{\partial
q_k}-H_{p_kq_j}\frac{\partial p_i}{\partial q_k}\right)+ \sum\sum
H_{p_kp_m} \left( \frac{\partial p_m}{\partial q_i}\frac{\partial
p_j }{\partial q_k}-\frac{\partial p_m}{\partial
q_j}\frac{\partial p_i}{\partial q_k}\right),
\]
we can write equation (\ref{derivada material de la vorticidad})
in the form
\begin{equation}\label{derivada material de la vorticidad vis}
    \frac{\partial }{\partial t}\left( \frac{\partial p_i}{\partial q_j}-\frac{\partial p_j}{\partial
    q_i}\right)+\sum\dot q_k\frac{\partial }{\partial q_k}\left( \frac{\partial p_i}{\partial q_j}-\frac{\partial p_j}{\partial
    q_i}\right)=
\end{equation}
\[
\sum H_{p_kq_i}\left(\frac{\partial p_j}{\partial
q_k}-\frac{\partial p_k}{\partial q_j}\right)+
H_{p_kq_j}\left(\frac{\partial p_k}{\partial q_i}-\frac{\partial
p_i}{\partial q_k}\right)+\]\[\sum\sum H_{p_kp_m} \left(
\frac{\partial p_m}{\partial q_i}\frac{\partial p_j }{\partial
q_k}-\frac{\partial p_m}{\partial q_j}\frac{\partial p_i}{\partial
q_k}\right).
\]

Now we assume that the mass matrix is diagonal, so that
\begin{equation}\label{condicion de la matriz de masas}
    \sum\sum H_{p_kp_m} \left(
\frac{\partial p_m}{\partial q_i}\frac{\partial p_j }{\partial
q_k}-\frac{\partial p_m}{\partial q_j}\frac{\partial p_i}{\partial
q_k}\right)=\sum m_k \left( \frac{\partial p_k}{\partial
q_i}\frac{\partial p_j }{\partial q_k}-\frac{\partial
p_k}{\partial q_j}\frac{\partial p_i}{\partial q_k}\right)=
\end{equation}
\[
\sum m_k \left( \frac{\partial p_k}{\partial q_i}\frac{\partial
p_j }{\partial q_k}-\frac{\partial p_k}{\partial
q_j}\frac{\partial p_i}{\partial q_k}\right)=
\]
\[
\sum m_k \left( \frac{\partial p_k}{\partial q_i}\frac{\partial
p_j }{\partial q_k}-\frac{\partial p_i}{\partial
q_k}\frac{\partial p_j }{\partial q_k}+\frac{\partial
p_i}{\partial q_k}\frac{\partial p_j }{\partial q_k} -
\frac{\partial p_k}{\partial q_j}\frac{\partial p_i}{\partial
q_k}\right)=\]
\[
\sum m_k \left( (\frac{\partial p_k}{\partial q_i}-\frac{\partial
p_i}{\partial q_k})\frac{\partial p_j }{\partial
q_k}+(\frac{\partial p_j }{\partial q_k} - \frac{\partial
p_k}{\partial q_j})\frac{\partial p_i}{\partial q_k}\right).
\]

Equation (\ref{derivada material de la vorticidad vis}) gives us
the \emph{material derivative} of the vorticity of the field
$p_i$. From equation (\ref{condicion de la matriz de masas}) we
can see that, under this assumption---that the matrix of masses is
diagonal---, if the vorticity of $(p_1,\cdots,p_n)$ is zero at
$t=t_0$, it will be zero at any other time. Therefore, in those
cases, there is a function $\Phi_i$, defined in space-time such
that:
\begin{equation}\label{integral de la accion}
    p_i=\frac{\partial \Phi}{\partial q_i},\ H=-\frac{\partial \Phi}{\partial
    t}.
\end{equation}

Obviously, $\Phi$ is a solution of the Hamilton Jacobi equation:
\begin{equation}\label{Hamilton-Jacobi}
    H(q_1,\cdots,q_n,\frac{\partial \Phi}{\partial q_1},\cdots,\frac{\partial \Phi}{\partial
    q_n},t)+\frac{\partial \Phi}{\partial t}=0.
\end{equation}

Here we are assuming that the trajectories of the particles in the
ensemble do not cross each other, which cannot be granted from the
sole condition on the vorticity. Therefore we can expect that the
solution of (\ref{Hamilton-Jacobi}) will have singularities,
depending on the initial conditions.

But suppose that $(p_1,\cdots,p_n)$ is not a potential field. The
vorticity of $(-p_1,\cdots,-p_n,H)$ will not be zero and,
according to our previous conclusions, there is at least  one
field $(-A_1,\cdots,-A_n,\Theta)$ such that:
\begin{equation}\label{introduccion del potencial vectorial}
    \frac{\partial \Theta}{\partial t}-\sum \frac{\partial A_i}{\partial
    q_i}=0,
\end{equation}
and $(-p_1-A_1,\cdots,-p_n-A_n,H+\Theta)$ is the best
approximation of $(-p_1,\cdots,-p_n,H)$ as a potential field.
Consequently, there is at least one function $\Phi$ such that
\begin{equation}\label{introduccion de la accion aparente}
    p_i=\frac{\partial \Phi}{\partial q_i}-A_i, H=-\frac{\partial \Phi}{\partial
    t}-\Theta.
\end{equation}
The function $\Phi$ is a solution of the differential equation
\begin{equation}\label{Hamilton-Jacobi con potencial vector}
    H(q_1,\cdots,q_n,\frac{\partial \Phi}{\partial q_1}-A_1,\cdots,\frac{\partial \Phi}{\partial
    q_n}-A_n,t)+\Theta+\frac{\partial \Phi}{\partial t}=0,
\end{equation}
which is the equation we had obtained, had we started with a
potential field $(p_1,\cdots,p_n)$ and another Hamilton's
function:
\begin{equation}\label{funcion de hamilton alternativa}
    H'(q_1,\cdots,q_n,p_1,\cdots,p_n,t)=H(q_1,\cdots,q_n,p_1-A_1,\cdots,p_n-A_n,t)+\Theta
\end{equation}

It is clear that equation (\ref{Hamilton-Jacobi con potencial
vector}) is not enough to determine the fields $\Phi$ and
$\vec{A}$, but this difficulty is easily surmounted. To show this
in a way that is free of mathematical complexities, we will
consider the case of a single particle, where the Hamilton's
function of the original system is:
\begin{equation}\label{hamiltoniano de una particula}
    H(\vec{p},\vec{q})=\frac{\vec{p}^2}{2m}+V.
\end{equation}

The new Hamilton's function is
\begin{equation}\label{hamiltoniano de una particula con vorticidad}
   H'(\vec{q},\vec{p})=\frac{(\vec{p}-\vec{A})^2}{2m}+V+\Theta,
\end{equation}
which is formally analogous to the Hamilton's function of a
particle in an electromagnetic field. The corresponding
Hamilton-Jacobi equation is:
\begin{equation}\label{equacion de hamilton jacobi modificada para una particula}
    H'(\vec{q},\grad \Phi)+\frac{\partial \Phi}{\partial t}=0.
\end{equation}

From (\ref{hamiltoniano de una particula con vorticidad}) we see
that the function $\Theta$ and the vector $\vec{A}$ have to be
chosen so that the corresponding Lorentz force is equal to zero
for the given field:
\begin{equation}\label{condicion adicional en le caso de una particula}
    -\grad \Theta - \frac{\partial \vec{A}}{\partial t} + \frac{\partial H'}{\partial \vec{p}}(\vec{q},\grad
    \Phi)\times(\grad\times \vec{A})=\vec{0}.
\end{equation}

Equations (\ref{equacion de hamilton jacobi modificada para una
particula}) and (\ref{condicion adicional en le caso de una
particula}) constitute the generalization of Hamilton-Jacobi
theory to the consideration of fields with vorticity.

To make an example, we have considered an ensemble of free
particles that satisfy the initial conditions:
\begin{equation}\label{condiciones iniciales ejemplo}
    \vec{v}(\vec{r}_0,t_0)=\vec{\omega}\times\vec{r}_0,
\end{equation}
where $\vec{\omega}=\omega\hat{k}$ is a constant vector in the
direction of the $z$ axis. The position at time $t$ of the
particle that occupies position $\vec{r}_0$ at time $t_0$ is
\begin{equation}\label{posicion al tiempo t ejemplo}
    \vec{r}(t)=\vec{r}_0+(\vec{\omega}\times{r}_0)(t-t_0).
\end{equation}
Or, in open form:
\begin{equation}\label{coordenada x ejemplo}
    \left(%
\begin{array}{c}
  x(t) \\
  y(t) \\
  z(t) \\
\end{array}%
\right)=\left(%
\begin{array}{ccc}
  1 & -\omega(t-t_0) & 0 \\
  \omega(t-t_0) & 1 & 0 \\
  0 & 0 & 1 \\
\end{array}%
\right)\left(%
\begin{array}{c}
  x_0 \\
  y_0 \\
  z_0 \\
\end{array}%
\right).
\end{equation}
The determinant of this system is greater than zero. Therefore,
the transformation is invertible. Actually, we have this:
\begin{equation}\label{inversion del flujo}
    x_0(x,y,z,t)=\frac{x+\omega(t-t_0) y}{1+\omega^2(t-t_0)^2},
\end{equation}
\[
   y_0(x,y,z,t)=\frac{y-\omega(t-t_0) x}{1+\omega^2(t-t_0)^2},
\]
\[
z_0(x,y,z,t)=z.
\]

From this  and (\ref{posicion al tiempo t ejemplo}) we can get the
field of momenta at time t:
\begin{equation}\label{campo de momento ejemplo}
  p_x = -m\omega y_0=\frac{-m\omega y + m\omega^2(t-t_0)x}{1+\omega^2(t-t_0)^2}
\end{equation}
 \[ p_y=m\omega x_0=\frac{m\omega x+ m\omega^2(t-t_0)
 y}{1+\omega^2(t-t_0)^2}\]
 \[p_z=0,\]
and the field of kinetic energy
\begin{equation}\label{energia cinetica ejemplo}
    K=\frac{1}{2}\frac{m\omega^2(x^2+y^2)}{1+\omega^2(t-t_0)^2}.
\end{equation}

The potential and solenoidal parts of the field of momenta are
clearly separated.
\begin{equation}\label{descomposicion ejemplo}
    \vec{p}=\grad \Phi-\vec{A},
\end{equation}
where
\begin{equation}\label{potencial del momento ejemplo}
    \Phi=\frac{1}{2}\frac{m\omega^2(t-t_0)(x^2+y^2)}{1+\omega^2(t-t_0)^2},
\end{equation}
\begin{equation}\label{parte rotacional del momento ejemplo}
  -A_x = \frac{-m\omega y }{1+\omega^2(t-t_0)^2},
\end{equation}
 \[ -A_y=\frac{m\omega x}{1+\omega^2(t-t_0)^2},\]
 and
 \[-A_z=0.\]
Also:

\begin{equation}\label{potencial teta para el ejemplo}
    \Theta=-K-\frac{\partial \Phi}{\partial
    t}=-\frac{m\omega^2(x^2+y^2)}{1+\omega^2(t-t_0)^2}.
\end{equation}

At this moment we can see that the alternative Hamilton's function
for this problem is:
\begin{equation}\label{funcion de hamilton alternativa para el ejemplo}
    H(\vec{r},\vec{p},t)=\frac{1}{2m}\left[\left(p_x-\frac{m\omega y}{1+\omega^2(t-t_0)^2}\right)^2%
    +\left(p_y+\frac{m\omega
    x}{1+\omega^2(t-t_0)^2}\right)^2+p_z^2\right]
\end{equation}
\[
-\frac{m\omega^2(x^2+y^2)}{1+\omega^2(t-t_0)^2}.
\]

Because of the way we have constructed this function, it is not
difficult to see that (\ref{potencial del momento ejemplo}) is a
particular solution of the Hamilton-Jacobi Equation
\begin{equation}\label{ecuacion de hamilton jacobi para el ejemplo}
\frac{1}{2m}\left[\left(\frac{\partial \Phi}{\partial x}-\frac{m\omega y}{1+\omega^2(t-t_0)^2}\right)^2%
    +\left(\frac{\partial \Phi}{\partial y}+\frac{m\omega
    x}{1+\omega^2(t-t_0)^2}\right)^2+\left(\frac{\partial \Phi}{\partial z}\right)^2\right]%
\end{equation}
\[
-\frac{m\omega^2(x^2+y^2)}{1+\omega^2(t-t_0)^2}+\frac{\partial
\Phi}{\partial t}=0.
\]

That (\ref{condicion adicional en le caso de una particula}) is
also satisfied can be easily proved from the equalities:
\begin{equation}\label{fuerza electrica ejemplo}
    -\grad \Theta -\frac{\partial A}{\partial
    t}=\frac{2m\omega^2(x\hat{i}+y\hat{j})}{1+\omega^2(t-t_0)^2}-%
    \frac{2m\omega^3(t-t_0)(y\hat{i}-x\hat{j})}{(1+\omega^2(t-t_0)^2)^2},
\end{equation}
\begin{equation}\label{velocidad en el ejemplo}
    \vec{v}=\frac{m\omega^2(t-t_0)x-\omega y}{1+\omega^2(t-t_0)^2)}\hat{i}+%
    \frac{\omega x + m \omega^2(t-t_0)^2y}{1+\omega^2(t-t_0)^2)}\hat{j}
\end{equation}
and
\begin{equation}\label{campo magnetico ejemplo}
   \grad\times\vec{A}=\frac{-2m\omega\hat{k}}{1+\omega^2(t-t_0)^2}.
\end{equation}

Passing to special relativity, in previous papers \cite{CHAVOYAI}
and \cite{CHAVOYAII}, we have observed that in the case of a field
of four-velocities that represents the motion of an infinite
ensemble of particles, the derivative of the four-velocity along
the corresponding world-lines
\begin{equation}\label{derivada de la cuatrovelocidad}
    \frac{d u_i}{ds}=u^j\frac{\partial u_i}{\partial x^j}
\end{equation}
is determined by its vorticity. This is so because, from the
condition $u^ju_j=1$, it follows that
\[
u^j\frac{\partial u_j}{\partial x^i}=0.
\]

Combining this and (\ref{derivada de la cuatrovelocidad}) we get
\begin{equation}\label{derivada de la cuatrovelocidad viz}
    \frac{d u_i}{ds}=u^j(\frac{\partial u_i}{\partial x^j}-\frac{\partial u_j}{\partial
    x_i}).
\end{equation}

Then---we see now---the vorticity of a field of four-velocities
cannot be zero at least it describes an ensemble of free
particles. If the particles move under the action of an
electromagnetic field we have:
\begin{equation}\label{ecuacion relativista de lorentz}
    u^j(\frac{\partial P_j}{\partial x^i}-\frac{\partial P_i}{\partial
    x^j})=0,
\end{equation}
where
\begin{equation}\label{cuatro impulso}
    P_i=mcu_i-\frac{q}{c}A_i,
\end{equation}
$m$ is the rest mass of the particles, and $q$ is the
corresponding electric charge.

We can write the vector $P_i$ in the form

\begin{equation}\label{descomposicion del cuatroimpulso}
    P_i=-\frac{\partial S}{\partial x^i}+\omega_i,
\end{equation}
where, because of (\ref{ecuacion relativista de lorentz}):
\begin{equation}\label{nueva ecuacion de hamilton jacobi}
   u^j(\frac{\partial \omega_i}{\partial x^j} -\frac{\partial \omega_j}{\partial
   x^i})=0,
\end{equation}
and
\begin{equation}\label{divergencia de omega}
    \frac{\partial \omega_i}{\partial x_i}=0.
\end{equation}

From (\ref{cuatro impulso}), (\ref{descomposicion del
cuatroimpulso}), (\ref{nueva ecuacion de hamilton jacobi}),
(\ref{divergencia de omega}), we get a non-linear system of
differential equations:
\begin{equation}\label{ecuacion de normalizacion}
    \left(-\frac{\partial S}{\partial x_k}+\frac{q}{c}A^k+\omega^k\right)%
    \left(-\frac{\partial S}{\partial
    x^k}+\frac{q}{c}A_k+\omega_k\right)=m^2c^2,
\end{equation}
\begin{equation}\label{ecuacion de la vorticidad}
\left(-\frac{\partial S}{\partial
x_k}+\frac{q}{c}A^k+\omega^k\right)(\frac{\partial
\omega_i}{\partial x^k} -\frac{\partial \omega_k}{\partial
   x^i})=0.
\end{equation}
\begin{equation}\label{divergencia de omega repetida}
    \frac{\partial \omega_i}{\partial x_i}=0.
\end{equation}

These equations encompass the most reasonable generalization of
Hamilton-Jacobi theory to special relativity since, as we said
before, a potential field of four-velocities represents
necessarily a field of free particles. Vorticity plays a special
role in relativistic Hamilton-Jacobi theory, and there is not a
physical reason to believe that the only \emph{real} fields of
four-velocities are those for which there is a scalar function
$\phi$ such that \cite[p. 488-509]{HOLLAND}
\begin{equation}\label{ecuacion restringida}
    mcv^u-\frac{q}{c}A^u=\frac{\partial \phi}{\partial x_u}.
\end{equation}
Quite the contrary, this reformulation of Hamilton-Jacobi theory
allows us to prove that it is possible to interpret the wave
functions of Klein-Gordon theory as describing the motion of an
ensemble of particles under the action of the electromagnetic
field, alone, without quantum potentials or uninterpreted hidden
variables, where the number of particles is not locally conserved.
\section{The Klein-Gordon Field}
From the Klein-Gordon equation
\begin{equation}\label{Klein-Gordon}
    -\hbar^2 \frac{\partial^2 }{\partial x_{\mu}\partial
    x^{\mu}}\Psi-\frac{2i\hbar}{c}A^{\mu}\frac{\partial \Psi}{\partial
    x^{\mu}}+q^2 A^{\mu}A_{\mu}\Psi=m^2c^2\Psi,
\end{equation}
we can easily show that
\begin{equation}\label{continuity}
    \frac{\partial }{\partial x_{\mu}}\left( \frac{i\hbar}{2}(\Psi^\star\frac{\partial \Psi}{\partial x^{\mu}}%
    -\Psi\frac{\partial \Psi^\star}{\partial
    x^{\mu}})-\frac{q}{c}A_{\mu}\Psi^\star\Psi\right)=0.
\end{equation}
From this, using the Madelung substitution
\begin{equation}\label{madelung}
    \Psi=\sqrt{\rho}e^{(i/\hbar)S},
\end{equation}
we prove that
\begin{equation}\label{continuity}
    \frac{\partial }{\partial x_{\mu}}\left(%
    \rho(-\frac{\partial S}{\partial
    x^{\mu}}-\frac{q}{c}A_{\mu})\right)=0.
\end{equation}

This suggest that $\rho$ could be interpreted as a density of
particles, in the system of reference where they are at rest and
that
\begin{equation}\label{first part of fourvelocity}
    v_{\mu}=-\frac{1}{mc}\frac{\partial
    S}{\partial x^{\mu}}-\frac{q}{mc^2}A_{\mu},
\end{equation}
which is usually rejected on the grounds that $v_{\mu}$ is not
unitary time-like by definition. From the previous section of this
paper we see now that we can complete the representation if we
suppose that
\begin{equation}\label{ecuacion del cuatro impulso}
    mcv_{\mu}=-\frac{\partial
    S}{\partial x^{\mu}}-\frac{q}{c}A_{\mu}+\omega_{\mu},
\end{equation}
where
\begin{equation}\label{condicion de la vorticidad}
    \left( -\frac{\partial
    S}{\partial
    x_{\mu}}-\frac{q}{c}A^{\mu}+\omega^{\mu}\right)%
    (\frac{\partial \omega_{\nu}}{\partial x^{\mu}}-\frac{\partial \omega_{\mu}}{\partial x^{\nu}})
    =0,
\end{equation}
and
\begin{equation}\label{ecuacion de normalizacion}
    \left( -\frac{\partial
    S}{\partial x^{\mu}}-\frac{q}{c}A_{\mu}+\omega_{\mu}\right)%
    \left( -\frac{\partial
    S}{\partial
    x_{\mu}}-\frac{q}{c}A^{\mu}+\omega^{\mu}\right)=m^2c^2.
\end{equation}

The continuity equation (\ref{continuity}) is transformed into:
\begin{equation}\label{creacion de particulas}
    \frac{\partial j^{\mu}}{\partial x^{\mu}}=\frac{\partial \rho \omega^{\mu}}{\partial
    x^{\mu}},
\end{equation}
that shows that the vector $\omega_{\mu}$ as linked to local
production/anihilation processes. This is not too strange if we
consider that, after all, this four-vector represents a kind of
electromagnetic field. Equation (\ref{condicion de la vorticidad})
is the condition it does not produce a Lorentz' force. However,
its very existence, implies the appearance of the corresponding
conserved current.

From equations (\ref{Klein-Gordon}) and (\ref{ecuacion del cuatro
impulso}) we can show that
\begin{equation}\label{introduccion de la constante de planck}
    -2mcv^{\mu}\omega_{\mu}+\omega^{\mu}\omega_{\mu}=%
    \frac{\hbar^2}{\sqrt{\rho}}\frac{\partial^2}{\partial x^{\mu}x_{\mu}}\sqrt{\rho}
\end{equation}

Equations (\ref{ecuacion del cuatro impulso}) to
(\ref{introduccion de la constante de planck}) make up a complete
description of an ensemble of particles that move under the action
of the electromagnetic field, in such way that the number of
particles is not locally conserved. Equations (\ref{creacion de
particulas}) and (\ref{introduccion de la constante de planck})
are the only ones that include the density of particles and
Planck's constant $\hbar$, which is then interpreted as an
empirical parameter that determines the local rate of
creation/anihilation of matter. The reinterpretation of Dirac's
field is similar to the one we have exposed in \cite{CHAVOYAII}.

In the low speed limit, we replace the four-vector $\omega_i$ by
its best approximation in the sense we studied in the second
section. This approximation does not meet the requirements of
Lorentz invariance, but Lorentz invariance is not required in the
low speed limit. Then we can follow the same line of though we
followed in \cite{CHAVOYAI} and \cite{CHAVOYAII} to recover
Schr\"odinger's and Pauli's equations.

\end{document}